\documentclass[reprint,amsmath,amssymb,aps,pra]{revtex4-1}

\usepackage{graphicx}
\usepackage{dcolumn}
\usepackage{bm}
\usepackage{amsmath}
\usepackage{amsthm}
\usepackage{amssymb}
\usepackage{float}
\usepackage{hyperref}
\usepackage{color}
\usepackage{epstopdf}
\usepackage{cleveref}
\usepackage{braket}
\usepackage[svgnames]{xcolor}
\hypersetup{hidelinks,colorlinks=true,allcolors=DarkBlue}

\begin{document}
\preprint{APS/123-QED}
\title{Tomography of a multimode quantum black box}
\author{Ilya A. Fedorov$^{1,2}$}
\author{Aleksey K. Fedorov$^1$}
\author{Yury V. Kurochkin$^1$}
\author{A. I. Lvovsky$^{1,2,3,4,}$}\email{LVOV@ucalgary.ca}
\affiliation
{
\mbox{$^{1}$Russian Quantum Center, 100 Novaya Street, Skolkovo, Moscow 143025, Russia} \\
\mbox{$^{2}$P. N. Lebedev Physics Institute, Leninskiy prospect 53, Moscow 119991, Russia}\\
\mbox{$^{3}$Institute for Quantum Science and Technology, University of Calgary, Calgary AB T2N 1N4, Canada}\\
\mbox{$^{4}$Canadian Institute for Advanced Research, 180 Dundas Street West, Toronto, Ontario M5G 1Z8, Canada}
}

\date{\today}

\begin{abstract}

We report a technique for experimental characterization of an $M$-mode quantum optical process, generalizing the single-mode coherent-state quantum-process tomography method 
[M.\,Lobino et al., {\href{http://dx.doi.org/10.1103/PhysRevA.90.043616}{Science {\bf 322}, 563 (2008)}}; A.\,Anis and A.I.\,Lvovsky, {\href{http://dx.doi.org/10.1103/PhysRevA.90.043616}{New J. Phys. {\bf 14}, 105021 (2012)}}].
By measuring effect of the process on multi-mode coherent states via balanced homodyne tomography, we obtain the process tensor in the Fock basis.
This rank-$4M$ tensor, which predicts the effect of the process on an arbitrary density matrix, is iteratively reconstructed directly from the experimental data via the maximum-likelihood method.
We demonstrate the capabilities of our method using the example of a beam splitter, reconstructing its process tensor within the subspace spanned by the first three Fock states. 
In spite of using purely classical probe states, we recover quantum properties of this optical element, in particular the Hong-Ou-Mandel effect.

\begin{description}
\item[PACS numbers]
03.65.Wj, 03.65.Ta, 42.50.Xa
\end{description}
\end{abstract}

\maketitle

\section{Introduction}

Precise understanding of the performance of individual quantum systems is a key requirement for the development of compound devices, {\it e.g.} quantum computers or secure communication networks.
This requirement gives rise to the problem of experimentally characterizing quantum systems as `black boxes: 
learning to predict their effect on arbitrary quantum states by measuring their effect on a limited number of ``probe" states. 
The art of solving this problem is referred to as quantum process tomography (QPT).

A straightforward approach to QPT consists of measuring the action of the black box on a set of states whose density operators form a spanning set in the space of all operators over a particular Hilbert space. 
Because any quantum process is a linear map with respect to density operators, this information is sufficient to fully characterize the process \cite{Poyatos1997}. 
However, such a direct method typically requires a large set of difficult-to-prepare probe states, and is consequently restricted to systems of very low dimension. 
Another possibility is the ancilla-assisted method \cite{D'Ariano2001} utilizing an input state that is a part of a fully entangled state in a larger Hilbert space. 
Although in this case only a single input is necessary thanks to the Jamiolkowski isomorphism \cite{Jamiolkowski1972}, both preparation of this state and tomography of the output state is, again, complicated, which dramatically limits the practicality of the method.

In application to optics, the coherent-state quantum process tomography (csQPT) \cite{Lobino2008, Lobino2009, Keshari2011} offers a practical solution. 
While being a member of the direct methods family described above, this technique uses only coherent states $|\alpha\rangle$ for probing the process $\mathcal{E}$, 
relying on the fact that these states span the space of operators over the optical Hilbert space (the optical equivalence theorem) \cite{Glauber1963b,Sudarshan1963}. 
The prediction for the output $\mathcal{E}\left(\hat\rho \right)$ of the black box in response to to an arbitrary input state $\hat\rho$ then involves integration of the measured output states $\mathcal{E}\left(|\alpha\rangle\langle\alpha| \right)$, 
weighted by corresponding Glauber-Sudarshan function $P_{\hat\rho}(\alpha)$, over the phase space.
A similar coherent-state based approach can also be used for the tomography of quantum measurements \cite{Lundeen2009, Zhang2012}. 
While being a case of the direct method described above, csQPT is relatively easy to implement in an experiment, since coherent states are readily obtained from lasers, and their amplitudes and phases are easy to control.

On the other hand, $P_{\hat\rho}(\alpha)$ is a generalized function, typically highly singular. 
Therefore the process reconstruction involving that function may either suffer from inaccuracies or involve an unreasonably large number of required probe states. 
Moreover, the procedures proposed in Refs.~\cite{Lobino2008,Keshari2011} evaluate each element of the process tensor individually, and can hence lead to unphysical (non-trace preserving or non-positive) process tensors.

The above shortcomings are absent in a method known as MaxLik csQPT, which exploits the Jamiolkowski isomorphism to reduce the QPT problem to the well-studied problem of the quantum state estimation, 
and applies the likelihood maximization technique to estimate the process tensor \cite{Hradil2004}. 
In this way, one can perform the reconstruction without leaving the physically plausible space. 
MaxLik csQPT has been proposed in Ref.~\cite{Anis2012} and successfully realized for nondeterministic singe-mode processes \cite{Kumar2013, Cooper2014}.

In this work, we expand csQPT beyond the ``single input --- single output'' case, which covers only a few of practically relevant quantum optical black boxes. 
The need for our study is dictated by the growing fields of quantum optical communication and logic, which are impossible without multimode processing. 
Examples include multimode quantum memories \cite{Gisin2009,Moiseev2010} and logic gates for processing photonic qubits \cite{Monroe,Pittman2003}, to name a few.
Although our experiment is in the optical domain, the theory and methodology of csQPT can be employed on a much broader scale. 
It applies to any physical system whose Hamiltonian is equivalent to that of the harmonic oscillator --- such as superconducting cavities, atomic spin ensembles and nanomechanical systems. 
In all of these, coherent states are the simplest to prepare and are hence most suitable as probe states in QPT. 

\section{Multimode MaxLik csQPT}

Our method generalizes the single-mode MaxLik csQPT approach \cite{Anis2012}, which we briefly outline below. 
We work in the Fock basis and represent a general $M$-mode quantum process $\mathcal{E}$ by a tensor of rank $4M$ which maps the input density matrix into the output one:
\begin{equation}\label{eq1}
	\rho^{out}_{\underline{j},\underline{k}} = \langle{\underline{j}|\mathcal{E}(\rho^{in})|\underline{k}}\rangle = \sum_{\underline{n},\underline{m}}{\mathcal{E}^{\underline{n},\underline{m}}_{\underline{j},\underline{k}}\rho^{in}_{\underline{n},\underline{m}}},
\end{equation}
where underlined symbols $|\underline{i}\rangle=|i_1,\dots,i_M\rangle$ refer to multimode Fock states.
In practice, the infinite dimensions of both input $\mathcal{H}$ and output $\mathcal{K}$ optical Hilbert spaces are truncated to the $N+1$ lowest Fock states, so that $i_k \in 0\dots N$.

In the experiment, the black box is tested with a set of $M$-mode coherent probe states $|{\underline\alpha}\rangle = |\alpha_1,\dots,\alpha_K\rangle$.
For every probe state, the output channels are examined by homodyne measurements, 
which gives a set of quadrature data $\{\underline X_i, \underline\theta_i\}$, where ${\underline\theta}_i=(\theta_{i1},\dots,\theta_{iM})$ is the set of local oscillator (LO) phases associated with the $i$th measurement.

To provide enough information about the process, the probe states should cover the volume of interest in the multimode phase space corresponding to the energies up to the chosen photon truncation number $N$. 
Because the mean quadrature variance of the $N$-photon state equals $N+1/2$, this volume corresponds to a  $2M$-dimensional hypersphere of radius $\sqrt{N+1/2}$. 
On the other hand, a single multimode set of coherent states corresponds to a  hypersphere of radius $\sqrt{1/2}$. 
Therefore the number of the necessary probe states can be estimated as $(2N+1)^M$. 

Our  process reconstruction method relies on the Jamiolkowski isomorphism, relates the superoperator $\mathcal{E}$ to an operator $\hat{E}$ on the product of $\mathcal{H}$ and $\mathcal{K}$ spaces:
\begin{equation}
	\hat{E} = \sum_{\underline{n},\underline{m},\underline{j},\underline{k}}{\mathcal{E}^{\underline{n},\underline{m}}_{\underline{j},\underline{k}} \ket{\underline{n}} \bra{\underline{m}} \otimes \ket{\underline{j}} \bra{\underline{k}}}.
\end{equation}
In this way, the process reconstruction is reduced to a more familiar problem of state reconstruction. 
The physicality of the process $\mathcal{E}$ requires it to be completely positive and trace preserving. 
These conditions are equivalent to the requirement that the corresponding Jamiolkowski operator be positive semidefinite and that $\mathrm{Tr}_\mathcal{K} [\hat{E}] = \hat{I}_\mathcal{H}$, where $\hat{I}$ is identity operator. 
The latter condition is readily extended to conditional (trace-reducing) processes as discussed in Refs.~\cite{Anis2012,Kumar2013}.

The maximum likelihood reconstruction consists of finding an operator $\hat{E}$ which maximizes the probability of obtaining the harvested data set $\{\underline X_i, \underline\theta_i\}$. 
Mathematically, this is equivalent to maximization of the functional
\begin{equation}\label{eq2}
	\mathcal{L}(\hat{E}) = \sum\limits_{i,j} \ln p(\alpha_j, i) - \mathrm{Tr}[\hat{\Lambda}\hat{E}],
\end{equation}
where $\hat{\Lambda}$ is Hermitian matrix of Lagrange multipliers incorporating the trace-preservation condition, and
\begin{equation}
\begin{split}
	p(\alpha, i)&=\mathrm{Tr}\left[\mathcal{E} (\ket{\alpha}\bra{\alpha}) \hat{\Pi}_{{\underline\theta}_i}(\underline X_i)\right] = \\
	&=\mathrm{Tr}\left[\hat{E} \ket{\alpha}\bra{\alpha} \otimes \hat{\Pi}_{{\underline\theta}_i}(\underline X_i) \right]
\end{split}
\end{equation}
is probability of registering $i$th outcome for the probe state $\ket{\alpha}$ and $\hat{\Pi}_{\underline\theta_i}(\underline X_i)=\ket{\underline X_i, \underline\theta_i}\bra{\underline X_i, \underline\theta_i}$ is the projector corresponding to the $i$th measurement outcome.
For deterministic processes, operator $\hat{E}$ maximizing the likelihood (\ref{eq2}) satisfies the extremal condition \cite{Anis2012, Hradil2004}
\begin{equation}\label{eq5}
	\hat{E} = \hat{\Lambda}^{-1} \hat{R} \hat{E} \hat{R} \hat{\Lambda}^{-1},
\end{equation}
where
\begin{eqnarray}\label{eq6}
	\hat{R} = \sum\limits_{i,j} \frac{\ket{\alpha_j^*}\bra{\alpha_j^*}\otimes \hat{\Pi}_{\underline\theta_i}(\underline X_i)}{p(\alpha_j, i)}, \\
	\hat{\Lambda} = \left(\mathrm{Tr}_\mathcal{K} \left[\hat{R}\hat{E}\hat{R}\right] \right)^{1/2} \otimes \hat{\mathcal{I}}_\mathcal{K}.
	\label{eq7}
\end{eqnarray}
Equations (\ref{eq5})--(\ref{eq7}) can be solved iteratively, starting from an unbiased $\hat{E}^{(0)}{=}\hat{\mathcal{I}}_{\mathcal{H}\otimes\mathcal{K}}/\mathrm{dim} \mathcal{K}$. 
Due to the Hermitian nature of operators $\hat{R}$ and $\hat{\Lambda}$, $\hat{E}$ remains positive semidefinite at each iteration. 
Together with the trace preservation constraint, this assures physicality of the reconstructed process. 
The likelihood functional (\ref{eq2}) is convex over the space of positive semidefinite operators, which eliminates the possibility of the iteration process stopping at a local maximum.

\section{Tomography of beam splitting}

The process of choice for testing the capability of our method is beam splitting. 
Its paramount importance in quantum optics needs no proof: all linear optical devices (interferometers, waveguide couplers, loss channels, etc.) are equivalent to single beam splitters (BSs) or sets thereof. 
Any single BS was recently shown to be generator of universal linear optics \cite{Bouland2014}. 
Accompanied by single photon sources and photon detectors, BSs enable quantum computation \cite{Knill2001}. 
In some form, a BS is present in any imaginable optical setting. 
In addition, the BS Hamiltonian is paramount in interfacing quantum information between harmonic oscillator systems of different nature, 
e.g. between an electromagnetic mode and either an atomic ensemble \cite{Hammerer2010}, or an electromagnetic mode and a nanomechanical oscillator \cite{Aspelmeyer2013}.

Although the operation of the BS is consistent with classical physics (coherent state inputs lead to coherent state outputs, and vice versa), its response to nonclassical input gives rise to quantum phenomena. 
A striking example is the Hong-Ou-Mandel effect: when two photons impinge upon a symmetric BS, they appear only in pairs at one of its outputs~\cite{Hong1987}. 
Our technique reveals this quantum effect in spite of using only classical states in measurements.

\begin{figure}[t]
\includegraphics[width=3.4in]{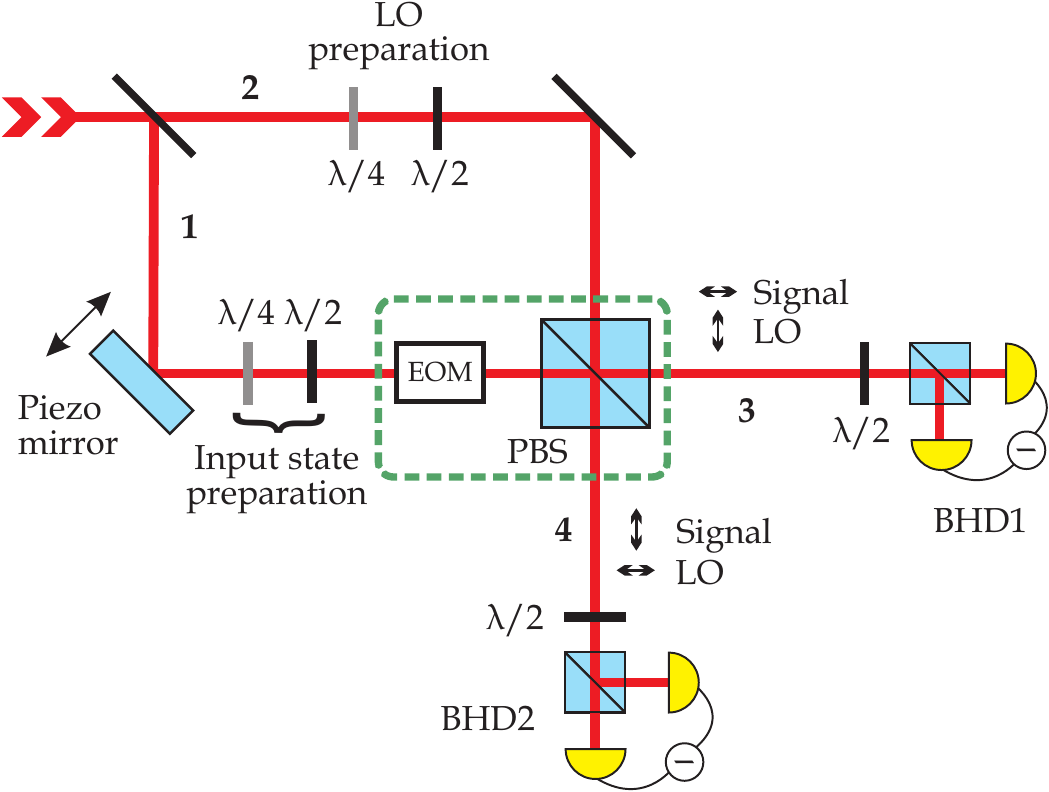}
\caption
{
Experimental setup. 
The BS process (encompassed by a green dashed line) is implemented in the polarization basis by an EOM to which a quarter-wave voltage is applied, and subsequent PBS. 
The input channels of the process are the horizontal and vertical modes of spatial mode 1; the output channels are the horizontal polarization of spatial mode 3 and the vertical polarization of spatial mode 4. 
The LOs for homodyne measurements are incident onto the PBS in the two polarization modes of spatial mode 2, thereafter emerging in  the vertical polarization of spatial mode 3 and the horizontal polarization of spatial mode 4.
}
\label{p1}
\end{figure}

The BS has previously been characterized by QPT in the role of a Bell-state filter \cite{Mitchell2003} and an amplitude damping channel \cite{Bongioanni2010}. 
In both these studies, tomography of the BS as a process on a multimode Hilbert space has been incomplete: limited to a specific photon number subspace of that space. 
Our technique is free of this shortcoming. 
It allows one to predict output of the process for any arbitrary Fock states and their superpositions in the input, up to a certain cut-off photon number.

Our technique is different from a recently developed methods for characterizing linear optical networks \cite{Rahimi-Keshari2013} and Gaussian processes \cite{Wang2013} in that it makes no assumptions about the content of the black box, 
in particular about its Gaussianity or linear-optical character. 
Although for the demonstration we do use a device which is both linear and Gaussian, our approach can be successfully applied to a multimode process of any nature.

The light source in our experiment is a mode-locked Ti:Sapphire laser (Coherent Mira 900), which emits pulses at $780$ nm with a repetition rate of 76 MHz and a pulse width of $\sim 1.8$ ps. 
In order to stabilize and control the relative phases of the inputs and outputs, we realize symmetric beam splitting with respect to the horizontal and vertical polarization modes in the same spatial channel, marked 1 in Fig.~\ref{p1}. 
The polarizations are mixed using an electrooptical modulator (EOM) with its optical axis oriented at $45^\circ$ to horizontal and a $\lambda/4$ voltage applied to it. 
A polarizing beam splitter (PBS) subsequently separates the output modes spatially for detection. 
Our black box is thus implemented by combination EOM + PBS. 
In the Heisenberg picture, this process has the  form
\begin{equation}\label{eq21}
	\left[ {\begin{array}{c}
	a_1^{out} \\
	a_2^{out} \\
	\end{array} } \right]
	=
	\frac{1}{2}
	\left[ {\begin{array}{cc}
	1 + i & 1 - i \\
	1 - i & 1 + i \\
	\end{array} } \right]
	\left[ {\begin{array}{c}
	a_1^{in} \\
	a_2^{in} \\
	\end{array} } \right],
\end{equation}
where $a_{1,2}^{in,out}$ are photon annihilation operators of the input and output modes. 
The relative amplitudes and phases of the input coherent states are set using a $\lambda/2 + \lambda/4$ waveplate pair.

The measurement of the output is performed using balanced homodyne detectors (BHDs) \cite{Kumar2012a} in both output channels. 
To this end, we introduce two LOs in orthogonal polarizations in spatial mode 2, so the central PBS directs them into the two output spatial channels (Fig.~\ref{p1}). 
In each output channel of the PBS, we then find the signal and LO in orthogonal polarizations. 
For homodyne detection, these polarizations are mixed in each channel using a combination of a $\lambda/2$ plate oriented at $22.5^\circ$ to the horizontal and an additional PBS. 
The relative phases the LOs can be controlled by two wave plates, while their common phase is slowly scanned using a piezo-mounted mirror in the signal channel.

\begin{figure*}[htbp]
\includegraphics[width=6.5in]{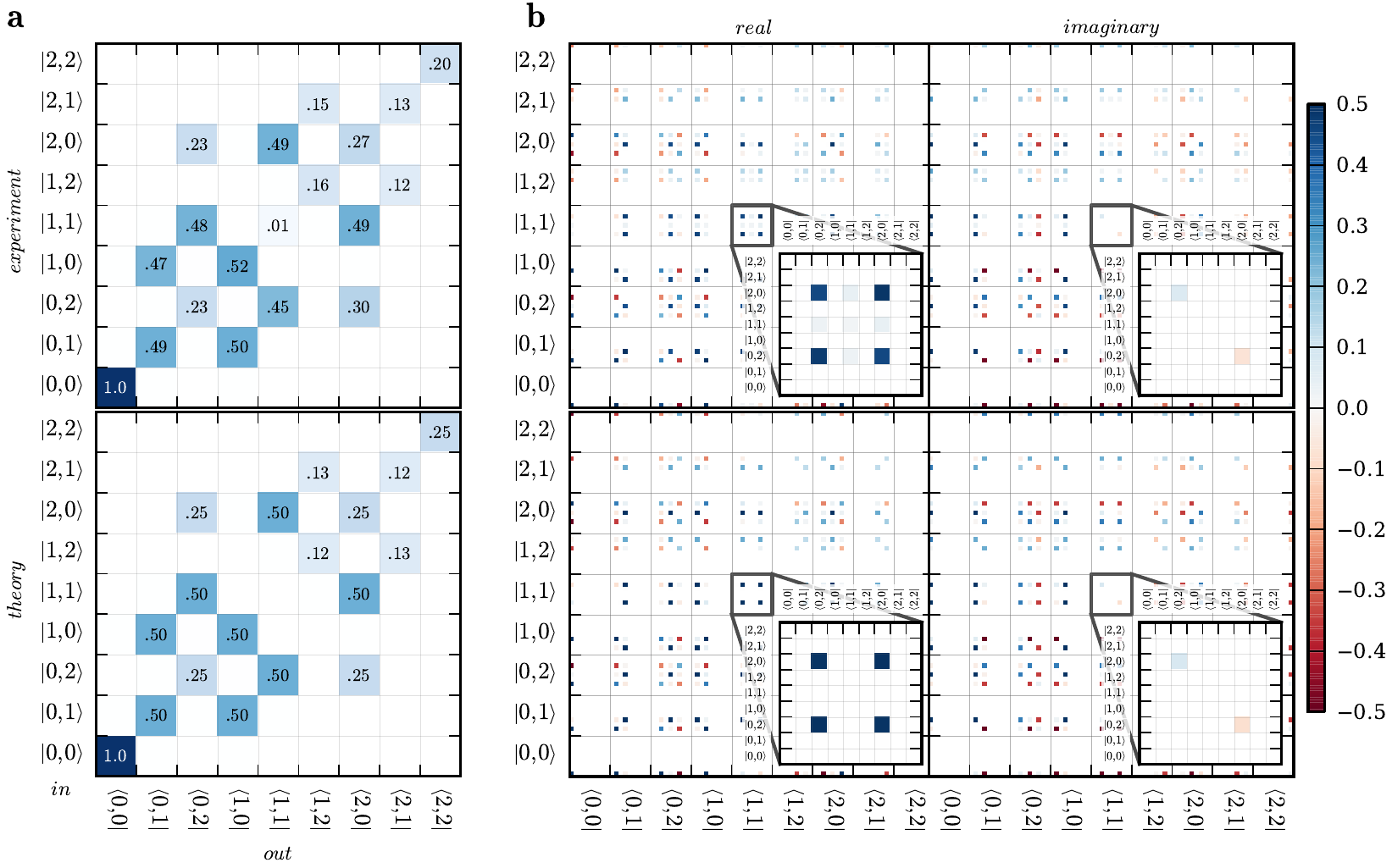}
\caption
{
Reconstructed (top) and theoretically expected (bottom) process tensor in the Fock space up to $N'=2$. 
a) Elements of the tensor corresponding to the diagonal elements of the input and output density matrices.  
Numbers give amplitudes of the non-zero cells.  
The element $\left| 1,1\right\rangle \rightarrow \left| 1,1\right\rangle$ corresponds to the coincidence probability in a Hong-Ou-Mandel measurement. 
b) The real (left) and imaginary (right) parts of the full tensor. Each large cell corresponds to a specific element of the input density matrix, while the content of each large cell gives the output density matrix. 
Insets show the magnified output for the input state $\left| 1,1\right\rangle$.
}
\label{p3}
\end{figure*}

\section{Evaluating the process tensor}

The process reconstruction is simplified by its invariance with respect to the global phase shift. 
That is, if both input phases are shifted by some phase $\theta$, so will be the output state. 
This invariance is a consequence of the symmetric nature of time: a global phase shift by $\theta$ is equivalent to a shift in time by $\theta/\omega$, where $\omega$ is the optical frequency. 
If the ``black box" is not connected to any external clock (such as in our case), it will respond to a signal that is shifted in time by the same amount in the output. 
The effect of phase invariance on the process tensor can be determined from the fact that a phase shift of both modes will transform density matrix elements according to
\begin{eqnarray}
	\rho^{in}_{n_1,n_2,m_1,m_2}&\to&\rho^{in}_{n_1,n_2,m_1,m_2}e^{i\theta(n_1 + n_2 - m_1 - m_2)}, \nonumber  \\
	\rho^{out}_{j_1,j_2,k_1,k_2}&\to&\rho^{out}_{j_1,j_2,k_1,k_2}e^{i\theta(j_1 + j_2 -k_1 - k_2 )}. \nonumber
\end{eqnarray}

Reconciling this with Eq.~(\ref{eq1}), we find that only elements such that $j_1 + j_2 -k_1 - k_2 = n_1 + n_2 - m_1 - m_2$ can be nonzero in tensor $\mathcal{E}^{\underline n,\underline m}_{\underline j,\underline k}$.

The process reconstruction requires knowledge of the LO phase vector $\underline \theta_{i}$, $i=1,\ldots,M$ at each moment in time both for the input and output of the black box. 
For a general phase-invariant process, this is equivalent to $2M-1$ unknown phase relations. 
This requirement makes a marked difference between the reconstruction of single-mode and multi-mode processes. 
In the single-mode case, many relevant processes exhibit intensity-independent phase behavior, which, in combination with the phase invariance, allows one to disregard phase relations between the input and output modes altogether. 
In multimode processes, however, this is almost always not the case: even in the relatively simple case of the present work, total phase control is essential for successful reconstruction. 

We acquire the phase vector $\underline \theta_{i}$ by periodically setting the EOM voltage to zero, so the black box becomes the identity process and the quadrature measurements correspond to the input states. 
This allows us to monitor all three required phase relations in real time. 
The inverse sine of the mean quadrature value for each set yields the differences $\underline\theta_{i}-\underline\theta_{in,i}$ of the LO and input state phases for both modes.

The switching between the BS and identity processes is performed with a period of 0.1 s, which is much faster than the characteristic time of phase fluctuations caused by air movements in the two interferometer channels. 
In this way, the evaluated LO phases can be translated to the process output measurements by taking into account the linear motion of the piezo.

We acquire a total of 48 sets of $10^6$ quadrature samples for three different relative phases of the LOs: $0.67$, $2.64$ and $5.29$ rad and, in addition to the vacuum, $16$ pairs of input coherent states, obtained by setting each waveplate at $0^\circ, 15^\circ, 30^\circ$ and $45^\circ$. 
Each pair of the input states has the same total energy corresponding to $0.9$ photons. This set of probe states is sufficient to reliably reconstruct the process up to a cut-off photon number of $2$.

We implement a two-step reconstruction process as prescribed by Ref.~\cite{Anis2012}. 
In the first step, we artificially inflate the reconstruction Hilbert space by choosing the cut-off point at $N=4$. 
This is necessary to ensure that both the input probe states and the output states are well accommodated in that space, which is required for the proper function of the reconstruction algorithm. 
However, the fraction of 3- and 4-photon terms in the Fock decomposition of the probe coherent states is relatively low, and so is their contribution to the log-likelihood functional.  
As a result,  the corresponding terms of the process tensor are not estimated accurately. 
To eliminate these inaccuracies, we truncate the reconstructed tensor to a lower maximum photon number $N'=2$ after the iterations have been completed \cite{Anis2012}.

The phase invariance property of the process kills about $90\%$ of $\approx 4\times 10^5$ tensor elements. 
The resulting dimensionality of the optimization space is close to that in the $8$-ion tomography done in work \cite{Haffner2005} and is computationally intensive. 
The iterative algorithm runs on an Intel Core i7 processor. Paralleled onto 4 of 8 computing cores, each iteration takes about 2 hours. 
The maximum-likelihood reconstruction algorithm appears to converge at around $100$ iterations.

\section{Results}

Fig.~\ref{p3} shows the result of the process reconstruction with $N'=2$ in comparison with the theoretical expectation according to Eq.~(\ref{eq21}) with an additional common phase delay of 0.8 rad. 
The elements of the process tensor associated with the diagonal elements of the input and output density matrices [Fig.~\ref{p3}(a)] have transparent physical meaning as probabilities of the corresponding transitions. 
In particular, the Hong-Ou-Mandel effect is represented by the probability of $\left| 1,1\right\rangle \rightarrow \left| 1,1\right\rangle$ transition, which is zero for ideally symmetrical BS and amounts to $0.01$ in the reconstructed tensor.

The data in Fig.~\ref{p3}(a) are only a small fraction of the full tensor shown in Fig.~\ref{p3}(b), which has $\sim 10^3$ non-zero, generally complex elements. 
These elements determine the phase behavior of the black box, and are equally important in the description of the process. 
The left and right columns of the grid present, respectively, the real and imaginary parts of the tensor, while the top and bottom rows correspond to the reconstruction result and the theoretical expectation. 
The insets in each panel shows the response of our black box to the Hong-Ou-Mandel query, the $\left| 1,1\right\rangle$ input state. 
The diagonal of the left (real) panel in Fig.~\ref{p3}(b) corresponds to the full panel in Fig.~\ref{p3}(a).

To characterize the quality of the reconstructed tensor shown in Fig.~\ref{p3}, we calculate the fidelity between the ideal and reconstructed processes in the Jamiolkowski state representation:
\begin{equation}\label{eq4}
	\mathcal{F} \left( \mathcal{E}, \mathcal{E}_{\rm est} \right) = \mathrm{Tr} \left[ \sqrt{ \sqrt{\mathcal{E}} \mathcal{E}_{\rm est} \sqrt{\mathcal{E}} }\right] = 0.95.
\end{equation}
We perform a few tests to find the source of this non-ideality. 
First, we quantify the physical imperfections of our black box by fitting the observed phase-dependent mean quadrature data by the theoretical prediction corresponding to an arbitrary BS. 
We obtain that the power transmittance corresponding to the best fit is 0.502. 
The fidelity between the processes associated with that slightly asymmetric BS and a symmetric one is 0.998, which shows that the physical errors (at least those which manifest in change of splitting ratio) are insignificant. 
Second, we evaluate the  statistical and systematic errors of the reconstruction using bootstrapping. 
Specifically, we simulate the quadrature data expected from a model BS and apply the MaxLik reconstruction algorithm to them to calculate a set of tensors $\mathcal{E}'_{{\rm est},i}$. 
The numbers of simulated data points, the dimensionality of the reconstruction space and the number of algorithm iterations were taken the same as in the real reconstruction procedure. 
We find $\mathcal{F} \left( \mathcal{E}, \mathcal{E}'_{{\rm est},i} \right)\sim 0.95$ for all $i$. 
Similar values are observed for the pairwise fidelities $\mathcal{F} \left( \mathcal{E}'_{{\rm est},i}, \mathcal{E}'_{{\rm est},j} \right)$ as well as for the fidelity $\mathcal{F} \left(\overline{ \mathcal{E}'_{{\rm est},i}}, \mathcal{E} \right)$ 
between the mean of the bootstrapping tensors and the theoretical one. 
These statistics show that the experimental fidelity of 0.95 results from both the inaccuracy of the numerical reconstruction algorithm and the statistical error conditioned by the limited amount of experimental data.

\section{Summary} 

We presented experimental csQPT reconstruction of the most common multimode optical process, the beam splitter. 
Our technique can be readily generalized to other processes, other physical systems and scaled up to a higher number of channels and larger state spaces thanks to the simplicity of the required optical measurements and probe state preparation.

\section*{Acknowledgment}

We thank Russian Quantum Center for support as well as A. Masalov, D. Mylnikov, A. Bozhenko, and S. Snigirev for helping with the experiment. 
AKF is a Fellow of the Dynasty foundation. 
AIL is a CIFAR Fellow.

\end{document}